\documentstyle[emulateapj,apjfonts]{article}
\newcommand{\lsim}{\mathrel{\mathop{\kern 0pt \rlap{\raise.2ex\hbox{$<$}}}
\lower.9ex\hbox{\kern-.190em $\sim$}}}
\newcommand{\gsim}{\mbox{\raisebox{-1.0ex}{$~\stackrel{\textstyle >}
{\textstyle \sim}~$ }}}
\newcommand{\be}{\begin{equation}}
\newcommand{\ee}{\end{equation}}
\newcommand{\ba}{\begin{eqnarray}}
\newcommand{\ea}{\end{eqnarray}}
\newcommand{\nn}{\nonumber}
\newcommand{\ep}{\epsilon}

\newcommand{\epet}{\tilde{\epsilon}_e}
\newcommand{\epe}{\epsilon_e}
\newcommand{\epp}{\epsilon_p}
\newcommand{\epn}{\epsilon_n}
\newcommand{\epmu}{\epsilon_{\mu}}

\newcommand{\nb}{n_{b}}
\newcommand{\nfe}{n_{f}^{e}}
\newcommand{\nfp}{n_{f}^{p}}
\newcommand{\gcm}{g \, cm^{-3}}
\newcommand{\me}{m_{e}}
\newcommand{\mpr}{m_{p}}
\newcommand{\mn}{m_{n}}
\newcommand{\mmu}{m_{\mu}}
\newcommand{\mpi}{m_{\pi}}
\newcommand{\gm}{\gamma}
\newcommand{\gme}{\gamma_{e}}
\newcommand{\gmp}{\gamma_{p}}
\newcommand{\gmn}{\gamma_{n}}
\newcommand{\gmu}{\gamma_{\mu}}

\newcommand{\xn}{x_{n}}

\newcommand{\xpi}{x_{\pi}}
\newcommand{\lame}{\lambda_{e}}
\newcommand{\lamp}{\lambda_{p}}
\newcommand{\lamn}{\lambda_{n}}

\newcommand{\ke}{\kappa_{e}}
\newcommand{\kmu}{\kappa_{\mu}}
\newcommand{\kp}{\kappa_{p}}
\newcommand{\kn}{\kappa_{n}}
\newcommand{\sze}{s_{z}^e}
\newcommand{\szp}{s_{z}^p}
\newcommand{\szn}{s_{z}^n}
\newcommand{\mpt}{\tilde{m}_{p}}
\newcommand{\mnt}{\tilde{m}_{n}}
\newcommand{\pz}{p_{\parallel}}
\newcommand{\pv}{p_{\perp}}
\lefthead{Suh \& Mathews}
\righthead{Cold ideal equation of state for strongly magnetized neutron-star matter}
\begin{document}
\title{Cold ideal equation of state for strongly magnetized neutron-star matter: 
effects on muon production and pion condensation}

\author{In-Saeng Suh$^{1}$ and G. J. Mathews$^{2}$}

\affil{Center for Astrophysics, Department of Physics, University of Notre Dame,
Notre Dame, Indiana 46556, USA \\
$^{1}$isuh@cygnus.phys.nd.edu; $\;\;$
$^{2}$gmathews@nd.edu}
\date{\today}
 
\begin{abstract}
Neutron stars with very strong surface magnetic fields have been
suggested as the site for the origin of observed soft gamma repeaters (SGRs).
In this paper we investigate the influence of such strong magnetic fields on
the properties and internal structure of these strongly magnetized neutron stars
(magnetars).
We study properties of a degenerate equilibrium ideal neutron-proton-electron
($npe$) gas with and without the effects of the anomalous nucleon magnetic moments
in a strong magnetic field.
The presence of a sufficiently strong magnetic field changes the ratio of
protons to neutrons as well as the neutron drip density.
We also study the appearance of muons as well as pion
condensation in strong magnetic fields. We discuss the possibility that boson
condensation in the interior of magnetars might be a source of SGRs.
\end{abstract}


\section{Introduction}

Among the more than two thousand observed cosmological gamma-ray bursts (GRBs),  
four recurrent sources, so-called soft gamma repeaters (SGRs), have been
identified and a fifth has probably been observed (\cite{hurley}).
They are believed to be a new class of $\gamma$-ray transients
separate from the source of classical GRBs.  
Observations of SGR 0526-66 (\cite{mazets}), SGR 1806-20 (\cite{murakami}),
and SGR 1900+14 (\cite{kouveliotou}) with RXTE, ASCA, and BeppoSAX have confirmed
the fact that these SGRs are newly born neutron stars that have very
large surface magnetic fields (up to $10^{15}$ G) based upon measurements of the  
spin-down timescale.
Recently, SGR 1627-41 has also been discovered by BATSE (\cite{woods}).
It is estimated that its magnetic field could be $B \gsim 5 \times 10^{14}$ G.  
The most recent source is SGR 1801-23 (\cite{cline}) observed by $Ulysses$, BATSE,
and KONUS-$Wind$. Such stars have been named magnetars (\cite{duncan,thompson95}).
[Note however that recently Harding et al. (1999) and Marsden et al. (1999) have
suggested that if relativistic wind outflow continuously dominates the spin-down of   
SGR 1806-20 and SGR 1900+14, then the surface dipole field may be too low   
to be consistent with a magnetar model.]

Magnetars have also been suggested as the site for anomalous X-ray pulsars (AXPs)   
(\cite{paradijs}) such as 1E 1841-045 (Kes 73) (\cite{gotthelf}), RX J0720.4-3125
(\cite{haberl}), and 1E 2259+586 (\cite{rho}). However, as another scenario, 
Chatterjee et al. (1999) proposed that these AXPs should have magnetic fields
similar to neutron stars inferred for ordinary radio pulsars and binary X-ray pulsars
in an accretion-powered X-ray emission model.

Whether or not magnetars are the source of SGRs or AXPs,
as relics of stellar interiors, the study of the magnetic fields in and around
degenerate stars should give important information on the role such fields play in
star formation and stellar evolution. Indeed, the origin and evolution of stellar
magnetic fields remains obscure.
As early as Ginzburg (1964) and Woltjer (1964) it was proposed that the magnetic
flux ($\Phi_B \sim BR^2$) of a star is conserved during its evolution and
subsequent collapse to form a remnant white dwarf or neutron star. A main sequence star
with radius $R \sim 10^{11}$ cm and surface magnetic field
$B \sim 10 - 10^4$ G [magnetic A-type stars have typical surface fields $\lsim 10^4$ G
(\cite{ST})] would thus collapse to form a white dwarf with $R \sim 10^9$ cm
and $B \sim 10^5 - 10^8$ G, or a neutron star with $R \sim 10^6$ cm and
$B \sim 10^{11} - 10^{14}$ G. Indeed, shortly after their discovery (\cite{hewish})
pulsars were identified as rotating neutron stars (\cite{gold}) with magnetic fields
$B \sim 10^{11} - 10^{13}$ G consistent with magnetic field amplification by flux
conservation.

Recently, Thompson and Duncan (1993) have invoked a convective dynamo mechanism  
to suggest that the magnetic dipole field of young neutron stars could realistically
reach values of the order of $10^{14} - 10^{15}$ G, i.e., $10^2 - 10^3$ times     
stronger than ordinary pulsars.
Moreover, the internal magnetic field of a star may not necessarily be reflected in its
surface magnetic field (\cite{ruderman,galloway}).  Therefore, the total strength of
internal magnetic fields remains unknown. Nevertheless, it is expected that appreciably
higher magnetic fields can exist in the interiors of neutron stars (\cite{ruderman}).

Ultimately, the allowed internal field strength of a star is constrained by
the scalar virial theorem (cf. \cite{chandra,ST}),
\[
2T + W + 3 \Pi + {\cal M} = 0,
\]
where $T$ is the total kinetic energy, $W$ is the gravitational potential energy,
$\Pi$ is the internal energy, and ${\cal M}$ is the magnetic energy.
For a star of size $R$ and mass $M$, this gives a maximum interior field strength of
$B \sim 2 \times 10^8 (M/M_{\odot}) (R/R_{\odot})^{-2}$  G.
For neutron stars with $R \approx 10$ km and $M \approx M_{\odot}$,
the maximum interior field strength could thus reach $B \lsim 10^{18}$ G (\cite{lerche}).
Numerical studies (\cite{bocquet}) have confirmed that neutron stars with ultrastrong
magnetic fields are stable up to the order of $10^{18}$ G.
They also have found that for such values the maximum mass of neutron stars increases by
13 - 29 \% relative to the maximum mass of non-magnetized neutron stars.
This is similar to the case of magnetic white dwarfs (\cite{SM20a}).

The strength of the internal magnetic field in a neutron star could, in principle, be
constrained by any observable consequences of a strong magnetic field. For example,
rapid motion of neutron stars may be due to anisotropic neutrino emission
induced by a strong magnetic field (e.g., see \cite{janka}). One could also consider  
the effect of magnetic fields on the thermal evolution (\cite{heyl,baiko}) and the
maximum mass (\cite{vshiv}) of neutron stars.
Recently, Chakrabarty et al. (1995) have investigated the gross properties of cold nuclear
matter in a strong magnetic field in the context of a relativistic Hartree model
and have applied their equation of state to obtain the maximum masses and radii for
magnetic neutron stars.

Since strong interior magnetic fields modify the nuclear equation of state for   
degenerate stars, their mass-radius relation will be changed relative to that of
nonmagnetic stars. Recently, we have obtained a revised mass-radius relation      
for magnetic white dwarfs (\cite{SM20a}) with the equation of state for electrons
in a strong magnetic field.
For strong internal magnetic fields of $B \sim 4.4 \times (10^{11} - 10^{13}$) G,   
we have found that both the mass and radius increase distinguishably and the mass-radius
relation of some observed magnetic white dwarfs may be better fit if strong internal fields
are assumed.

In this work, we now extend this study to an investigation of the effect of
magnetic fields on the internal properties of neutron stars as well.
If ultrastrong magnetic fields exist in the interior of neutron stars,
such fields will primarily affect the behavior of the residual charged particles.
Moreover, contributions from the anomalous magnetic moment (AMM) of the particles
in a strong magnetic field should also be significant (\cite{broderick}).
In particular, in a strong magnetic field, complete spin polarization of the neutrons
occurs due to the interaction of the neutron magnetic moment with the magnetic field.
Therefore, we consider both cases with and without the effects of the AMM.

Even so, standard internal properties such as the nuclear equation of state,
neutron drip, and the threshold density of new particles will be modified by a strong
magnetic field.
For purposes of illustration, we will consider a degenerate ideal noninteracting
neutron-proton-electron ($npe$) gas in equilibrium (\cite{ST}).
We find that under conditions of charge neutrality and chemical equilibrium,
the presence of a sufficiently strong magnetic field
changes the ratio of protons to neutrons as well as the threshold density for the  
appearance of muons and pion condensation.

In Sec. 2 we first review the properties of electrons and describe the equation of state
for a particle gas in an external magnetic field.
In Sec. 3 we first consider the case without the particle AMM and
we derive the proton to neutron ratio in an ideal $npe$ gas for the lowest Landau
level analytically. We also numerically obtain the neutron appearance density and proton
concentration, $Y_p$, in magnetic fields. In Sec. 4 we include an effect of the particle
AMM on the equation of state and the adiabatic index.
In Sec. 5 we calculate the muon threshold density in a magnetic field.
We consider pion production and condensation in strongly magnetized neutron-star matter
in Sec. 6. Finally, we discuss the possibility that boson condensation in the interior of
magnetars is a source of SGRs.

\section{Equation of state for particles in a magnetic field}

\subsection{Electrons and muons}

Let us begin by considering the properties of an electron in an external magnetic field
(\cite{landau,JL,canuto,schwinger}).
For a sufficiently high field, the energy states of an electron in a magnetic field are
quantized and its properties are modified accordingly.
The cyclotron energy becomes comparable to the electron rest-mass energy and the electrons
in the excited Landau levels become relativistic.

In order to investigate these effects for electrons we must first solve the Dirac equation
in an external, static, and homogeneous magnetic field.
With a uniform magnetic field $B$ along the $z$-axis and the convenient choice of gauge
for the vector potential, $i.e.$, $A_{0} = 0$ and $\vec{A} = (-yB, 0, 0)$,
we obtain the electron wavefunctions and energy dispersion in a magnetic
field (\cite{JL,schwinger}). The dispersion relation relation for an electron   
propagating through a magnetic field is
\be
E_{\nfe}^e = [p_{z}^{e \, 2} c^2+ \me^2 c^4 + 2 \hbar c e B \nfe]^{1/2} + \me c^2 \ke,
\label{electron1}
\ee
where $\nfe = n + {1 \over 2} + \sze$, in which
$n$ is the  principal quantum number of the Landau level,
$\sze =\pm 1/2$ is the electron spin projection onto the magnetic field direction,
$e$ is the electron charge, $c$ is the speed
of light, $\hbar$ is Planck's constant,
$p_{z}^{e}$ is the electron momentum along the $z$-axis, and
$\me$ is the rest mass of the electron.
Here, let us use a definition $\gme \equiv B/B_{c}^{e}$
where $B_{c}^{e} = \me^{2} c^3 /e \hbar =4.414 \times 10^{13}$ G.
In Eq. (\ref{electron1}), $\ke = - \; (\alpha/4 \pi) \gamma_e$, for $\gamma_e \ll 1$ and
$\ke = (\alpha/4 \pi) [{\rm ln}(2 \gamma_e) - (C + \frac{3}{2})]^2 + \cdots$,
for $\gme \gg 1$, where $\alpha \simeq 1/137$ and $C = 0.577$ is Euler's constant
(\cite{schwinger}).

The main modification of an electron in a magnetic field comes from the available
density of states for the electrons (\cite{landau}).
The electron state density in the absence of a magnetic field,
\be
\frac{2}{\hbar^3} \int \frac{d^{3}\vec{p}}{(2 \pi)^3},
\label{electron2}
\ee
is replaced with
\be
\frac{2}{\hbar^2 c} \sum_{n} \sum_{\sze}
\int {eB\over(2\pi)^2} dp_z
\label{electron3}
\ee
in a magnetic field.
This modification affects the thermodynamic properties of the electron gas.

Let us consider a gas of electrons at zero temperature in a magnetic field (\cite{blandford}).
From Eq. (1) we can define the electron Fermi energy $E_{F}^{e}$ for an arbitrary Landau
level $\nfe$ as
\be
E_{F}^{e} \equiv [\me^2 c^4 + p^{e \, 2}_{F} c^2 + 2 \hbar c e B \nfe ]^{1/2} +  \me c^2 \ke.
\label{electron4}
\ee
Here $p^{e}_F$ denotes the electron Fermi momentum.

Now we can obtain all of thermodynamic quantities in terms of the Fermi energy, Eq.
(\ref{electron4}),
and the phase space integration, Eq. (\ref{electron3}), in a magnetic field.
The number density of electrons in a magnetic field is then given by
\be
n_e  = \frac{\gme}{2 \pi^2} \left( \frac{\me c}{\hbar} \right)^3
\; \zeta_e (\epe),
\label{electron5}
\ee
where $\epe$ is defined as $\epe \equiv E_{F}^e / \me c^2$ and
\be
\zeta_e (\epe) = \sum_{n=0}^{n_{max}^e} \sum_{\sze} \;
\sqrt{(\epe - \ke)^2 - (1 + 2 \gme \nfe)}.
\label{electron6}
\ee
The maximum Landau level $n_{max}^e$ for a given electron Fermi energy $\epe$ and magnetic
field strength $\gme$ is given by
\be
n_{max}^e = int \Bigg[ \frac{(\epe-\ke)^2 - 1}{2 \gme}  - \Bigg(\frac{1}{2} + \sze \Bigg)
\Bigg] \ge n,   
\label{electron7}
\ee
where $int[x]$ means an integer value of the argument $x$.
The pressure of an ideal electron gas in a magnetic field is then
\be
P_e = \frac{\gme}{4 \pi^2} \me c^2 \left( \frac{\me c}{\hbar} \right)^3 \; \Phi_e (\epe),
\label{electron8}
\ee
where
\ba
\Phi_e (\epe) = \sum_{n = 0}^{n_{max}^e} \sum_{\sze}
\Bigg[ (\epe - \ke) \sqrt{(\epe - \ke)^2 - (1 + 2 \gme \nfe)} \nn \\
- (1 + 2 \gme \nfe) \; \mbox{ln}\Bigg(\frac{(\epe-\ke) + \sqrt{(\epe - \ke)^2 -
(1 + 2 \gme \nfe)}}{\sqrt{1 + 2 \gme \nfe}} \Bigg) \Bigg].
\ea
Similarly, the energy density is
 \be
{\cal E}_e (\epe) = \frac{\gme}{4 \pi^2} \me c^2 \left(\frac{\me c}{\hbar}\right)^3
\; \chi_e (\epe), 
\label{electron9}
\ee
where
\ba
\chi_e (\epe) = \frac{1}{2} \sum_{n=0}^{n_{max}^e} \sum_{\sze}
\Bigg[ (\epe - \ke) \sqrt{(\epe - \ke)^2 - (1 + 2 \gme \nfe)} \nn \\
+ (1 + 2 \gme \nfe) \; \mbox{ln}\Bigg( \frac{(\epe - \ke) + \sqrt{(\epe - \ke)^2 -
(1 + 2 \gme \nfe)}}{\sqrt{1 + 2 \gme \nfe}} \Bigg) \Bigg].
\ea
From these, we obtain the energy per electron
\be
E_e (\epe) = \me c^2  \; \frac{\chi_e (\epe)}{\zeta_e (\epe)}.
\label{electron10}
\ee
Note, that as $\gme$ goes to zero, Eqs. (\ref{electron6}) and (\ref{electron8}) -
(\ref{electron10}) recover exactly the usual non-magnetic equation of state for
electrons (\cite{SM20a}).

We can obtain similar quantities for muons simply by replacing the electron quantities
by the corresponding muon quantities (e.g., replace $\gme = B/B_{c}^{e}$ by
$\gmu = B/B_{c}^{\mu}$,where $B_{c}^{\mu} = \mmu^2 c^3 / e \hbar$).
The muon AMM $\kmu$ has nearly the same value as
$\ke$. The difference is only $\kmu - \ke \simeq 0.59 \times 10^{-5}$ (\cite{schwinger73})
[The experimental value for this difference is $0.63 \times 10^{-5}$ (\cite{grandy})].   

\subsection{Protons}

Although the proton mass is much greater than the electron mass,
magnetic effects on protons can be as important as those on electrons (\cite{lai}).
For instance, the proton pressure is always much smaller than the electron pressure
at low density. But, ignoring the influence of the magnetic field on protons would
lead to the unphysical result of proton pressure dominance at low density.
Therefore, whenever the magnetic field significantly affects the electrons,
it also affects the protons.
    
The energy dispersion relation for protons $E_{\nfp}^{p}$ for an arbitrary Landau level
in a magnetic field is:
\be
E_{\nfp}^p = \Bigg[ p_{z}^{p \, 2} c^2+ \mpr^2 c^4 \big[ \big\{1 +  2 \gmp n_{f}^{p} \big\}^{1/2}
- \szp \frac{\mu_{N}^p B}{\mpr c^2} \big]^2 \Bigg]^{1/2},
\label{proton1}
\ee
where
$\nfp =  n + \frac{1}{2} - \szp$, $n$ is the  principal quantum
number of the Landau level, $\szp =\pm \frac{1}{2}$ is the $z$ component of the proton spin,
$p^{p}_z$ is the proton momentum along the $z$-axis, and
$\mu_{N}^p = (e \hbar / m_p c) \kp$, with $\kp = 2.79$, is the proton anomalous magnetic
moment.

The proton number density in a magnetic field is :
\be
n_p  = \frac{\gmp}{2 \pi^2} \left( \frac{\mpr c}{\hbar} \right)^3
\; \zeta_p (\epp),
\label{proton2}
\ee
where $\epp \equiv E_{F}^p / \mpr c^2$ and
\be
\zeta_p (\epp) = \sum_{n=0}^{n_{max}^p} \; \sum_{\szp}
\sqrt{\epp^2 - \mpt^{2}},
\label{proton3}
\ee
\be
\mpt = \sqrt{1 + 2 \gmp \nfp} - \szp \kappa_p \gmp.
\label{proton4}
\ee
The maximum Landau level $n_{max}^p$ for a proton in a magnetic field is given by  
\be
n_{max}^p = int \Bigg[\frac{(\epp + \szp \kappa_p \gmp)^2 - 1}{2 \gmp}
- \Bigg(\frac{1}{2} - \szp \Bigg) \Bigg].
\label{proton5}
\ee
The pressure of a proton gas in a magnetic field is
\be
P_p = \frac{\gmp}{4 \pi^2} \mpr c^2 \left( \frac{\mpr c}{\hbar} \right)^3
\; \Phi_p (\epp),
\label{proton6}
\ee
where
\be
\Phi_p (\epp) = \sum_{n=0}^{n_{max}^p} \sum_{\szp}
\Bigg[ \epp \sqrt{\epp^2 - \mpt^2}
 - \mpt^2 \; \mbox{ln}\Bigg( \frac{\epp + \sqrt{\epp^2 - \mpt^2}}{\sqrt{\mpt}}
\Bigg) \Bigg].
\ee
The energy density is
\be
{\cal E}_p (\epp) = \frac{\gmp}{4 \pi^2} \mpr c^2 \left(\frac{\mpr c}{\hbar}\right)^3
\; \chi_p (\epp),
\ee
\label{proton7}   
where
\be
\chi_p (\epp) = \frac{1}{2} \sum_{n=0}^{n_{max}^p} \sum_{\szp}
\Bigg[ \epp \sqrt{\epp^2 - \mpt^2}
 + \mpt^2 \; \mbox{ln}\Bigg( \frac{\epp + \sqrt{\epp^2 - \mpt^2}}{\sqrt{\mpt}} \Bigg)
\Bigg].
\ee
   
\subsection{Neutrons}

A neutral Dirac fermion can interact with an external electromagnetic field by means of
the Pauli non-minimal coupling. Then the energy dispersion relation
$E_{F}^{n}$ for neutrons in a magnetic field is given by
\be
E_{F}^n = \big[\pz^2 c^2 + (\sqrt{\mn^2 c^4 + \pv^2 c^2} + \szn \mu_{N}^n B)^2 \big]^{1/2}
\label{neutron1}
\ee
where $\szn =\pm \frac{1}{2}$ is the neutron spin projection onto the magnetic field direction,
$\pz$ and $\pv$ are the components of the neutron momentum parallel and perpendicular to the
magnetic field, and  $\mu_{N}^n = (e \hbar/\mn c) \kn$, with $\kn=-1.91$, is the neutron
anomalous magnetic moment.

In the calculations of thermodynamic quantities, the phase-space integration separates
into two steps which involve integration over $\pz$ and $\pv$ (e.g., see \cite{broderick}).
Then the neutron number density in a magnetic field is given by
\be
n_n  = \frac{1}{2 \pi^2} \left( \frac{\mn c}{\hbar} \right)^3
\; \zeta_n (\epn),
\label{neutron2}
\ee
where
\ba
&~&\zeta_n (\epn)= \sum_{\szn} \frac{1}{3} \Bigg[[\epn^2 - \mnt^2]^{3/2} \nn \\
&+& \frac{3}{2} \szn \kn \gmn \Big[\mnt \sqrt{\epn^2 - \mnt^2} + \epn^2 \Big\{
{\rm arcsin}\Big(\frac{\mnt}{\epn}\Big) - \frac{\pi}{2}\Big\} \Big] \Bigg].
\label{neutron3}
\ea
In the above Eqs (\ref{neutron2}) and (\ref{neutron3}),
we have defined $\epn \equiv E_{F}^n / \mn c^2$,
$\gmn \equiv B/B_{c}^{n}, \; B_{c}^{n} = e \hbar / m_{n}^{2} c^3$,
and
\be
\mnt = 1 + \szn \kn \gmn.
\label{neutron4}
\ee
The pressure of a neutron gas in a magnetic field is
\be
P_n = \frac{1}{24 \pi^2} \mn c^2 \left( \frac{\mn c}{\hbar} \right)^3
\; \Phi_n (\epn),
\label{neutron5}
\ee
where
\be
\Phi_n (\epn) = \sum_{\szn} \frac{1}{2} \Big[ \Phi_{n}^{0}
+ \szn \kn \gmn \Phi_{n}^{\kappa} \Big].
\label{neutron6}
\ee
In equation (\ref{neutron6}),
\be
\Phi_{n}^{0} = (2 \epn^2 - 4 - \mnt^2 + 9 \kn^2 \gm_{n}^{2}) {\cal P}
+ (4 - \mnt^2 - 9 \kn^2 \gm_{n}^{2}) \mnt^2 {\cal Q},
\label{neutron7}
\ee
\ba
\Phi_{n}^{\kappa} &=& \frac{1}{3} \mnt \Bigg[ 2 {\cal P} + 4 \{\mnt^2 + \mnt
+ 6(\kn^2\gm_{n}^{2} - 1) \} {\cal Q} \Bigg] \nn \\
&~&+ \Bigg[ \frac{10}{3} \epn^2 + 8 (\kn^2 \gm_{n}^{2} - 1) \Bigg] {\cal R},
\label{neutron8}
\ea
where we defined the quantities:
\ba
{\cal P} &=& \epn \sqrt{\epn^2 - \mnt^2}, \;\;\;
{\cal Q} = {\rm ln} \Bigg[\frac{\epn + \sqrt{\epn^2 - \mnt^2}}{\mnt}\Bigg], \nn \\
&~& {\cal R} = \epn \Bigg( {\rm arcsin} \Big(\frac{\mnt}{\epn}\Big) - \frac{\pi}{2} \Bigg).
\ea
Finally the energy density of a neutron gas in a magnetic field is
\be
{\cal E}_n (\epn) = \frac{1}{8 \pi^2} \mn c^2 \left(\frac{\mn c}{\hbar}\right)^3
\; \chi_n (\epn),
\label{neutron9}
\ee
where
\be
\chi_n (\epn) = \sum_{\szn} \frac{1}{6} \Bigg[ \chi_{n}^{0}
+ \szn \kn \gmn \, \chi_{n}^{\kappa} \Bigg],
\label{neutron10}
\ee
and
\be  
\chi_{n}^{0} = 3 \big[ (2 \epn^2 - \mnt^2) {\cal P} - \mnt^4 {\cal Q} \big],
\;\;\;\;\;
\chi_{n}^{\kappa} = 4 \big[ \mnt {\cal P} + \mnt^3 {\cal Q} + 2 \epn^2 {\cal R}
\big].
\label{neutron11}
\ee

\section{Inverse $\beta$-decay and neutron appearance in a strong magnetic field:
without the anomalous magnetic moments}

Let us first consider the physics of an $npe$ gas in a strong magnetic field
without the anomalous magnetic moments of particles, i.e., $\kappa_j = 0 \, (j = e,p,n)$.
This illustrates the dominant physics at moderate magnetic field strength.
For illustration, consider a homogeneous gas of free neutrons, protons, and electrons in
$\beta$-equilibrium in a uniform magnetic field.
At low densities, the most energetically favorable nucleus is $^{56}$Fe which
is the endpoint of thermonuclear reactions. As the density increases above
$\sim 10^4$ ${\rm \gcm}$, electrons become unbound and relativistic.
At sufficiently high densities, $\rho \gsim 8 \times 10^6$ ${\rm \gcm}$, protons in nuclei
are converted into neutrons via inverse $\beta$-decay:
\be
e^{-} + p \longrightarrow n + \nu
\label{beta1}.
\ee
Since the neutrinos can escape, energy is transported away from the system.
Thus, the equation of state in the star will be modified mainly due to the inverse
$\beta$-decay. The reaction (\ref{beta1}) can proceed whenever the electron acquires
enough energy to exceed the mass difference between protons and neutrons,
$Q = m_n - m_p = 1.293$ MeV.
The transformation of protons into neutrons, reaction (\ref{beta1}), is effective
whenever the $\beta$-decay reaction;
\be
n \longrightarrow p + e + \bar{\nu}
\label{beta2}
\ee  
is slower than the rate of electron capture by protons.
Reaction (\ref{beta2}) is blocked if the density is high enough that
all energetically available electron energy levels in the Fermi sea are occupied.
Thus, there is a critical density for the onset of reaction (\ref{beta1}).

Similar to the field-free case, we can take into account the above processes
in an intense magnetic field.
Assuming that a mixture of free neutrons, protons, and electrons are in equilibrium, then
reaction (\ref{beta1}) implies
\be
\mu_e + \mu_p = \mu_n,
\label{beta3}
\ee
where $\mu_j \equiv E_{F}^{j}$ ($j=e,p,n,\mu,\pi$) is the chemical potential of the $j$th
particle. We have set the neutrino chemical potential $\mu_{\nu_e}$ to zero.
Let us now define
\be
x_j \equiv \frac{p_{F}^{j}}{m_{j} c}, \;\;\; \ep_j \equiv \frac{\mu_{j}}{m_{j} c^2},
\;\;\; {\rm and} \;\;\; \lambda_j \equiv \frac{\hbar}{m_j c},
\label{beta4}
\ee
where $\lambda_j$ is the Compton wavelength of the $j$th particle.
   
From chemical equilibrium, Eq. (\ref{beta3}), we have
\be
\me c^2 \epe + \mpr c^2 \epp = \mn c^2 \epn,
\label{beta5}
\ee
and charge neutrality gives
\be
n_e = n_p.
\label{beta6}
\ee
In order to determine the equilibrium composition and hence the equation of state,
the above equations should be solved simultaneously.

Consider now the minimum density at which neutrons first appear in a strong magnetic
field. This neutron appearance density is determined by setting $n_n = 0$,
or $\ep_n = 1 = (1 + \xn^2)^{1/2}$.
Since the protons at this density are nonrelativistic, i.e., $\epp^{\ast} \approx 1$,
we approximately obtain $\epe^{\ast} \simeq 2.53$ as the specific electron chemical
potential at which neutrons first appear according to Eq. (\ref{beta5})
(Hereafter an asterisk is used to denote a threshold value
for the appearance of new particles.) Therefore, we have $n_{max}^e = 0$ for electrons
if $\gme > (\epe^{\ast \, 2} - 1)/2 \simeq 2.7$.
That is, for $B > 2.7 B_{c}^e \simeq 1.2 \times 10^{14}$ G,
electrons reside in the lowest Landau level.
Substantially $n_{max}^p$ may not be zero for $2 \gamma_p \ll 1$.
But, in order to compare the two cases of higher-Landau-level and lowest-Landau-level
occupation for the charged particles,
we simply take $n_{max}^p = 0$ for protons.

Since $Q$ and $\me$ are both much less than $\mn$,
from Eqs. (\ref{beta5}) and (\ref{beta6}) we then obtain the proton-to-neutron
ratio analytically when we assume that electrons and protons are in the lowest Landau level.
Then
\be
\frac{n_p}{n_n} = \Bigg[\frac{(2 \mn Q + C_{n}^{2} \, n_{n}^{2/3})^2 - 4 \mn^2 \me^2}
{4 C_{p}^{2} \, n_{n}^{2} (\mn^2 + C_{n}^{2} \, n_{n}^{2/3})}\Bigg]^{1/2}
\label{beta7},
\ee
where $C_p = 2 \pi^2 \mpr \lamp^3 / \gmp$, and $C_{n}^{3} = 3 \pi^2 \mn^3 \lamn^3$.
\placefigure{fig1}
\begin{center}
\vspace*{0.5cm}
{\epsfxsize=6.5cm
\epsfbox{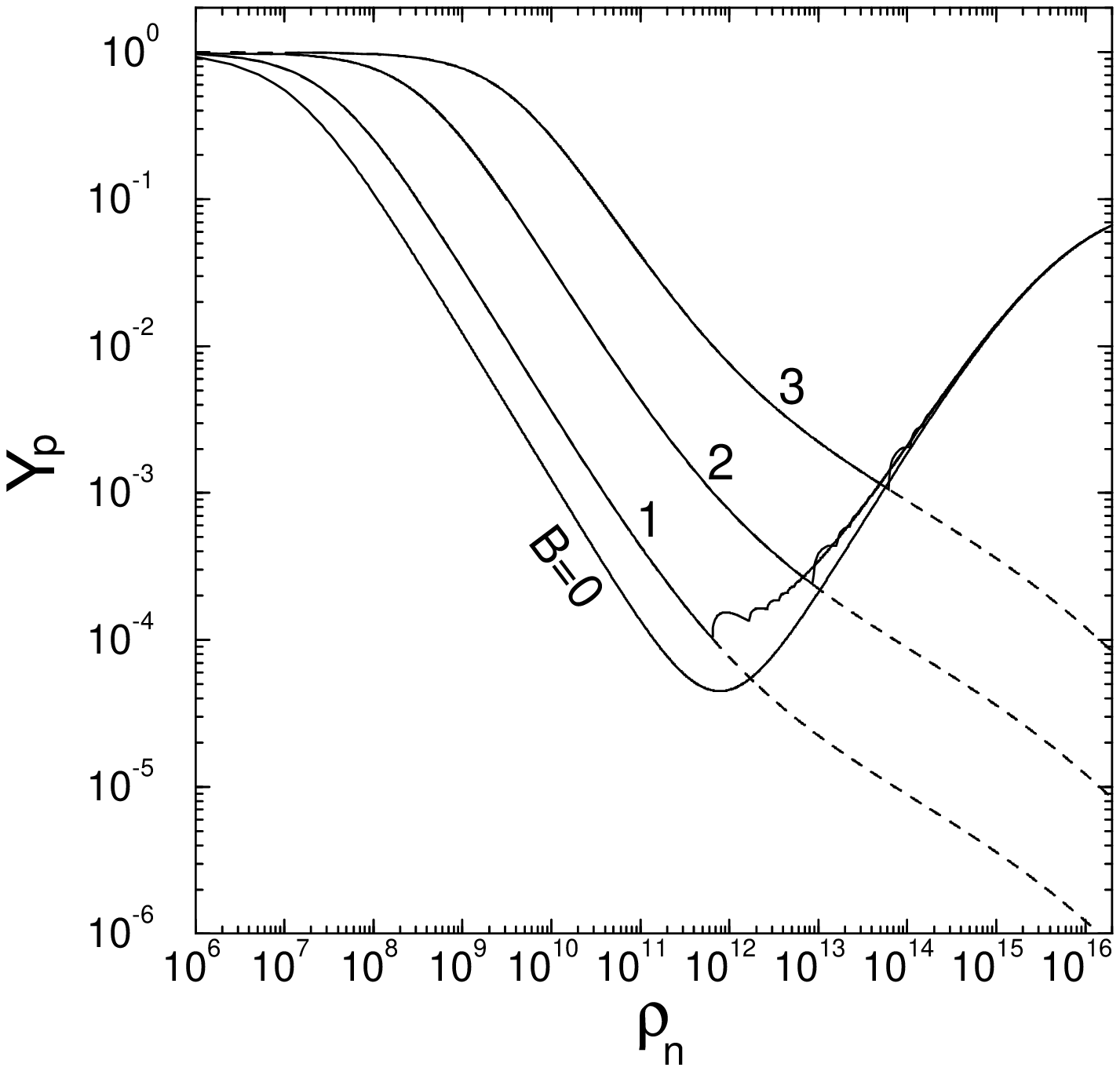}}
\vskip 0.2cm
\end{center}
{\small {\sc Fig. 1}
The proton fraction $Y_p = n_p / n_B$ without the AMM as a function of the neutron
density $\rho_n$ (in unit ${\rm \gcm}$) for the given value of log$\gme$,
$\gamma_e = B/B_{c}^{e}$'s.
The $B=0$ line is the non-magnetic case. The dashed lines occur if charged particles
are restricted in the lowest Landau level.
}
\vskip 0.5cm
Figure 1 shows the proton fraction $Y_p = n_p / n_B$ (where $n_B = n_n + n_p$)
as a function of the neutron density $\rho_n$ for values of $\gme$ less than $10^3$.
(For higher fields, we should take into account the nucleon AMM).
In the nonmagnetic case, the conditions of charge neutrality and chemical
equilibrium in a $npe$ gas lead to a threshold for an increase in the proton concentration
up to a value of $Y_p \simeq 1/9$ as $\rho_n$ exceeds $\approx 10^{12}$ ${\rm \gcm}$.
This means that the inverse $\beta$ decay is strongly suppressed by Pauli blocking in
neutron-rich nuclear matter which consists only of a $npe$ gas.
However, in the case of a strong magnetic field, if we assume that electrons
and protons are always in the lowest Landau level, then an increase in the concentration
of protons does not occur even as $\rho_n$ exceeds $10^{12}$ ${\rm \gcm}$.
That is, inverse $\beta$ decay is not suppressed in magnetic fields.   
Far from suppressing the inverse $\beta$ decay,
the magnetic field instead catalyzes the reaction. This means that rapid
neutron-star cooling can occur in a strong magnetic field through the direct URCA process
(\cite{baiko,leinson}).
[Note that Baiko \& Yakovlev (1999) have considered the direct URCA process
at the core of a neutron star ($\rho > 10^{14} \, {\rm \gcm}$) and not its crust.
In reality, the direct URCA process can never proceed at such low density as
$10^{12}$ ${\rm \gcm}$ because most protons are confined within the nuclei at
$\rho < 10^{14} \, {\rm \gcm}$ in realistic neutron-star matter].

However, electrons and protons, actually, are not in the lowest Landau
level for higher densities. Above a critical density, higher Landau levels begin
to contribute to the chemical potential of the electrons and protons and hence particle
number densities.
Ultimately, discrete Landau levels become continuous and the proton concentration
$Y_p$ reverts to the nonmagnetic limit as the density increases. As a result,
inverse $\beta$ decay can still be suppressed at high densities in strong magnetic fields.
Therefore, neutron-star rapid cooling may not be affected by the direct URCA process
even though it is enhanced in strong magnetic fields.
However, in order to enhance the cooling by the direct URCA process,
one can invoke other mechanisms such as boson condensation (Tsuruta 1998),
nucleon superfluidity (Yakovlev et al 1999), etc., if they exist.

The proton-to-neutron ratio, Eq. (\ref{beta7}), gives the number density
at which neutrons first appear:
\be
n_{n}^{\ast}(B) = n_{p}^{\ast}(B) = n_{e}^{\ast}(B) = \frac{\gme}{2 \pi^2 \lame^3}  
\Bigg[\frac{Q^2 - \me^2}{\me^2}\Bigg]^{1/2} .
\label{beta8}
\ee
Comparing with the zero-field result
\be
n_{n}^{\ast}(0) = \frac{1}{3 \pi^2}\frac{1}{\lame^3}
\Bigg[\frac{Q^2 - \me^2}{\me^2}\Bigg]^{3/2},
\label{beta9}
\ee
we obtain the relative density at which neutrons appear in a strongly magnetized neutron
star for $B > 2.7 B_{c}^e$ (e.g., see \cite{lai}):
\be
\frac{\rho_{n}^{\ast}(B)}{\rho_{n}^{\ast}(0)}
= \frac{3}{2} \gme \Bigg[\frac{\me^2}{Q^2 - \me^2} \Bigg]
= 0.277 \gme .
\label{beta10}
\ee
We can see that the neutron appearance density increases linearly with the magnetic
field $B$. This result is equivalent to one directly calculated from the general
form (\cite{ST}),
\ba
\rho_{n}^{\ast} (B) &\simeq& \mpr n_{e}^{\ast} (B)  \nn \\
&=& \mpr \frac{\gme}{2 \pi^2 \lame^3} \sum_{n = 0}^{n_{max}^e} \sum_{\sze}
\sqrt{(\epe^{\ast}-\ke)^2 - (1 + 2 \gme \nfe)},
\label{beta11}
\ea
with $\epe^{\ast} \simeq Q / \me $.

\section{Effects of anomalous magnetic moments}

Since $n_n = 0$ at neutron appearance, we have $\epn^{\ast} = 1 + \szn \kn \gmn$.
Then, from chemical equilibrium and charge neutrality, we obtain the electron Fermi energy
at neutron appearance $\epet^{\ast}$ when the nucleon AMM is included:
\be
\epet^{\ast} = \frac{\eta^2 - \mpr^2 + \me^2}{2 \me \eta},
\label{amm1}
\ee
where $\eta = \mn (1 + \szn \kn \gmn) + \mpr \szp \kp \gmp$. 
Note that $\eta$ depends on both $\szn$ and $\szp$.
Of course Eq. (\ref{amm1}) goes to $\epe^{\ast} \simeq Q/\me$ when $\kappa_{p, n} = 0$.
In Eq. (\ref{amm1}) $\eta$ is similar to the neutron effective mass in a magnetic field 
because $\eta$ becomes $\mn$ when $\kappa_{p, n} = 0$. 
Hence, for $\szp = -1/2$, it is possible for $\eta$ to have 
a negative value as $\gme$ increases. 
But this case is unphysical because $\epet^{\ast}$ becomes negative.
Therefore, we take the proton spin $\szp = + 1/2$ in this work.
Now we can obtain the neutron appearance density when the nucleon AMM is included 
\be
\frac{\rho_{n}^{\ast} (B)}{\rho_{n}^{\ast} (0)} = \frac{3}{2} 
\frac{\gme}{(\epe^{\ast \, 2} - 1)^{3/2}} 
\sum_{n=0}^{n_{max}^{e \ast}} \sum_{\sze}
\sqrt{(\epet^{\ast} - \ke)^2 - (1 + 2 \gme \nfe)}.
\label{amm2}
\ee
Here, $n_{max}^{e \ast}$ denotes the maximum electron Landau level at neutron
appearance.
The neutron threshold density $\rho_{n}^{\ast} (B)$
is plotted in figure 2 as a function of $\gme = B / B_{c}^{e}$.
We find that $\rho_{n}^{\ast}$ is significantly affected by the nucleon AMM
above log$\gme \approx 3$. Also, above this field strength, the neutron appearance
density is split according to the orientation of the neutron spin.
\placefigure{fig2}
\begin{center}
\vspace*{0.5cm}
{\epsfxsize=6.5cm
\epsfbox{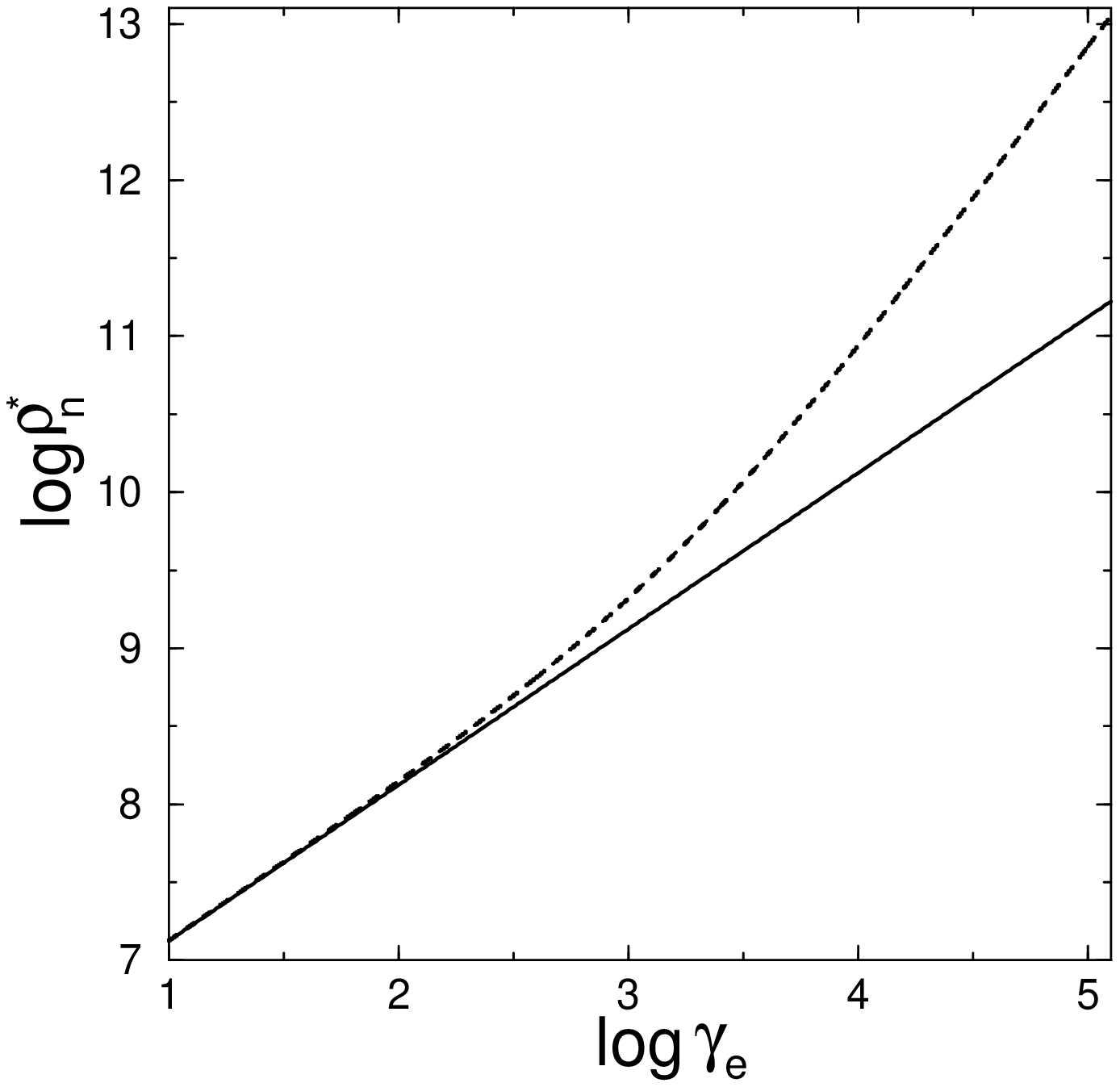}}
\vskip 0.2cm
\end{center}
{\small {\sc Fig. 2}
Neutron appearance density $\rho_{n}^{\ast}$ (in ${\rm \gcm}$) 
as a function of $\gme$.
The solid line is the case without the AMM. The dashed line corresponds to
$\szn = -1/2$ when the nucleon AMM's are included.
}
\vskip 0.5cm
Now the equation of state for a $npe$ gas at $\rho > \rho_{n}^{\ast} (B)$ can be determined
in terms of the parameter $\epe$.
The charge neutrality condition, Eq. (\ref{neutron11}), at a fixed $\gme$,
\be
\sum_{n = 0}^{n_{max}^e} \sum_{\sze} p_{F}^{e} (\epe, \sze)
=\sum_{n = 0}^{n_{max}^p} \sum_{\szp}  p_{F}^{p} (\epp, \szp)
\label{amm3}
\ee
gives $\epp$ for a given $\epe$.
We also obtain $\epn$ from the chemical equilibrium, Eq. (\ref{neutron10}).  
Note that both sides of Eq. (\ref{amm3}) have the same statistical weight for the excited 
Landau level when the AMM of the proton is ignored. However, the statistical weight of protons 
is changed when the proton AMM is included.

Finally, we obtain the total baryon density $n_B = n_p + n_n$ and  the proton
concentration $Y_p = n_p / n_B$. 
In figure 3 we can see that the nucleon AMM above a field strength of 
$\gme \gsim 10^3$ significantly affects the value of $Y_p$. 
The difference between the AMM and non-AMM results 
above a density of $\gsim 10^{12}$ ${\rm \gcm}$ comes from the polarization of the 
proton spin. Besides, the equation of state, the mass-energy density
$\rho = ({\cal E}_e + {\cal E}_p + {\cal E}_n)/c^2$ and
the pressure $P = P_e + P_p + P_n$, are straightforwardly determined.
In calculations of the equation of state and the adiabatic index, we take $\szn = -1/2$.
\placefigure{fig3}
\begin{center}
\vspace*{0.5cm}
{\epsfxsize=6.5cm
\epsfbox{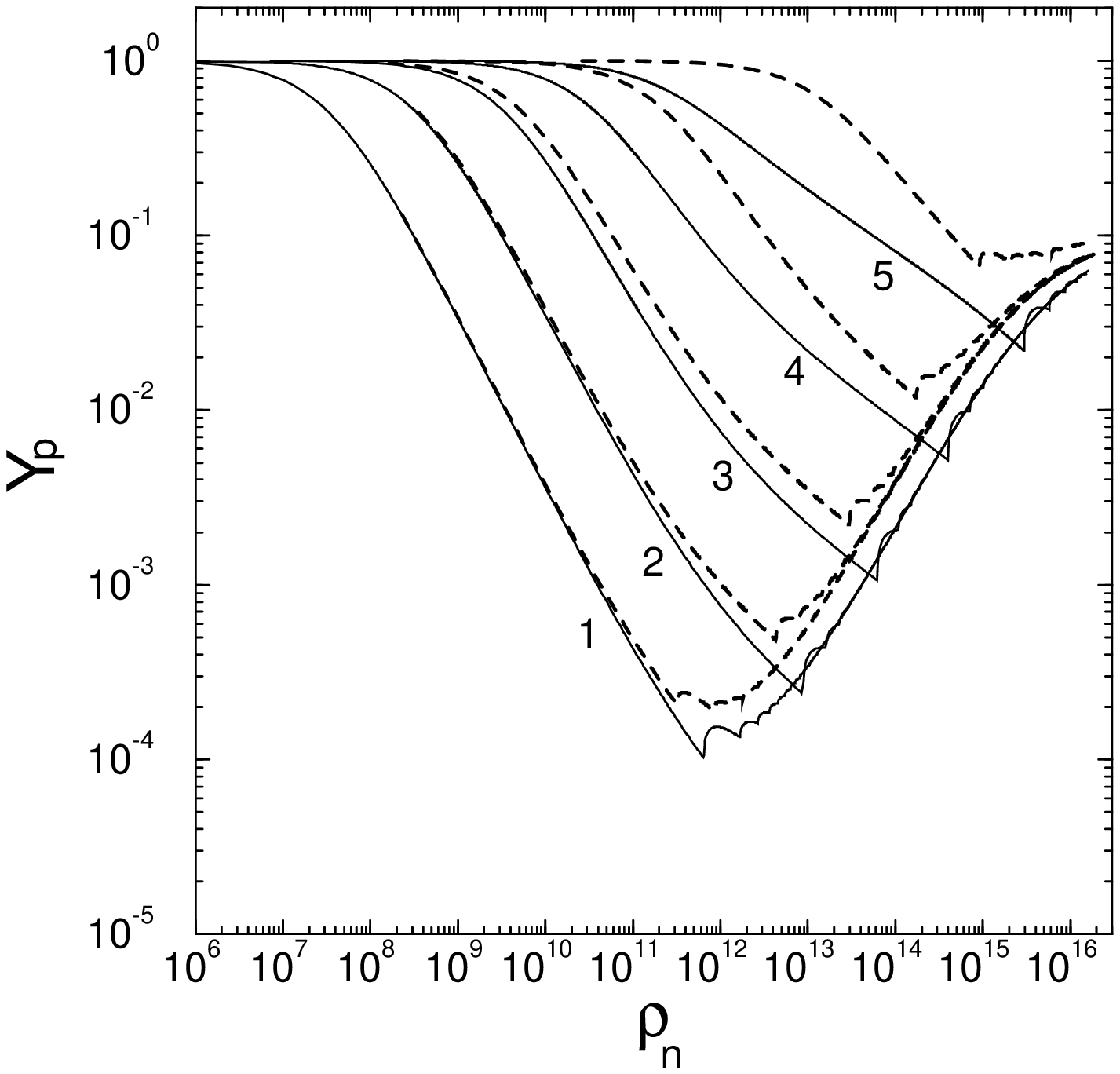}}
\vskip 0.2cm
\end{center}
{\small {\sc Fig. 3}
The proton fraction $Y_p = n_p / n_B$ with the AMM included as a function of the neutron
density $\rho_n$ (in unit ${\rm \gcm}$) for the given value of log$\gme$.
The solid line is for the case without the nucleon AMM.
The dashed lines are the cases of $\szn = -1/2$.
}
\vskip 0.5cm
Lai \& Shapiro (1991) have shown the equation of state and the adiabatic index
for a field strength less than log$\gme$ = 2.
In this work, we calculate those for a field strength less than log$\gme$ = 5.
Figure 4 shows the equation of state for a strongly magnetized $npe$ gas.
At low density ($\rho < \rho_{n}^{\ast}$) and high magnetic
field ($B \ge 4.4 \times 10^{14}$ G), we utilize the fact that electrons are
nonrelativistic and are in the lowest Landau level as in Lai \& Shapiro (1991).
We plot the adiabatic index $\Gamma \equiv {\rm d \, ln P / d \, ln}\rho$
as a function of $\rho$ in Figure 5. We find a non-zero $\Gamma$ at the 
neutron appearance as the magnetic field increases.
   
The adiabatic index is a crucial factor for understanding the global radial stability
of a star as well as the local sound speed (\cite{ST}).
In figure 5 we know that at neutron appearance $\Gamma$ has a non-zero value and the 
minimum value of $\Gamma$ increases as the magnetic field strength increases. 
But the nucleon AMM reduces the minimum of $\Gamma$ at neutron appearance.
A strange feature in figure 5 is that for a magnetic field strength greater
than $\sim 10^{15}$ G (log$\gme \gsim 2)$, an oscillatory behavior in the adiabatic
index begins to appear above a density of $\rho \gsim 10^{12}$ ${\rm \gcm}$.
This means that at high densities and fields $\Gamma$ is significantly affected by
the proton fraction even though it is small ($Y_p \lsim 0.1$).
Notice that for log$\gme \leq 2$, as the density increases above the neutron
appearance density $\rho_{n}^{\ast}$, the oscillatory behavior vanishes since neutron
pressure dominates.

This oscillatory behavior has the same physical origin as the well-known 
de Haas-van Alphen effect (\cite{landau}). It arises as electrons begin to fill the next
unoccupied Landau level. Increasing the density increases the occupation of this level
and does not lead to a rapidly increased pressure.
It would be interesting to understand the physical consequences of the oscillatory
behavior of the adiabatic index in strongly magnetized neutron stars. 
We speculate that this might drive a pulsational instability not unlike classical cepheids.
That is, as a region develops low $\Gamma$ it may be unstable to collapse.
At high density, however, $\Gamma$ suddenly stiffens and the region bounces as 
the $\Gamma$ is restored to a higher value.
To explore this possibility, however, it would be necessary to calculate the 
evolution of the interior of strongly magnetized neutron stars 
in a realistic magnetohydrodynamical model. This will be the subject of a future study
(\cite{SM20b}).
\placefigure{fig4}
\begin{center}
\vspace*{0.5cm}
{\epsfxsize=6.5cm
\epsfbox{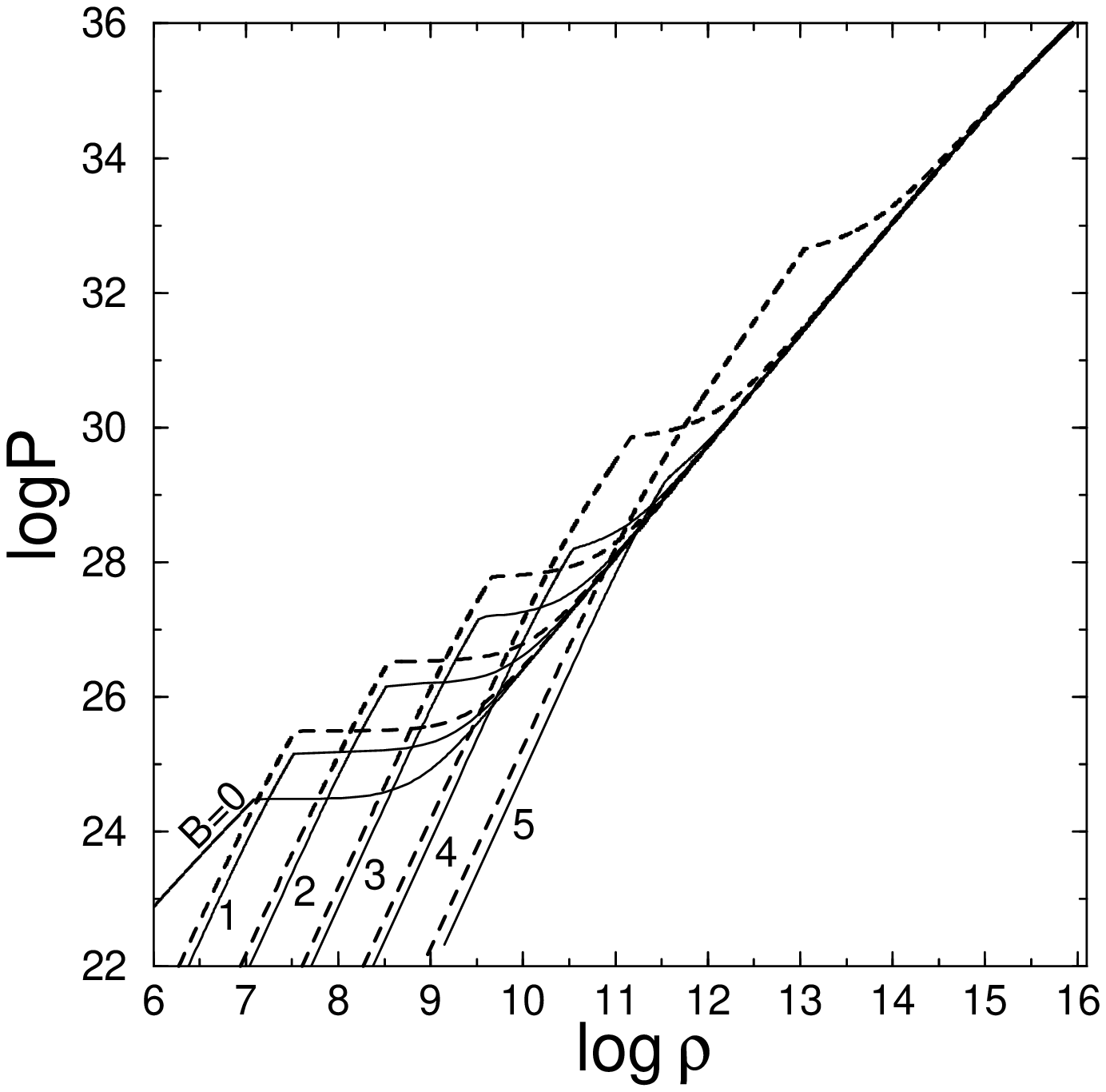}}
\vskip 0.2cm
\end{center}
{\small {\sc Fig. 4}
Pressure $P$ (in dyn/${\rm cm^2}$) vs. total density $\rho$ (in ${\rm \gcm}$) for a $npe$ gas
in magnetic fields of various strengths, log$\gme$.
The $B=0$ line corresponds to the non-magnetic case.
The solid line are for the case without the AMM.
The dashed lines are for $\szn = -1/2$ with the AMM included.
Numbers are labeled as in Fig. 3.
}

\vskip 0.5cm

\placefigure{fig5}
\begin{center}
\vspace*{0.5cm}
{\epsfxsize=6.5cm
\epsfbox{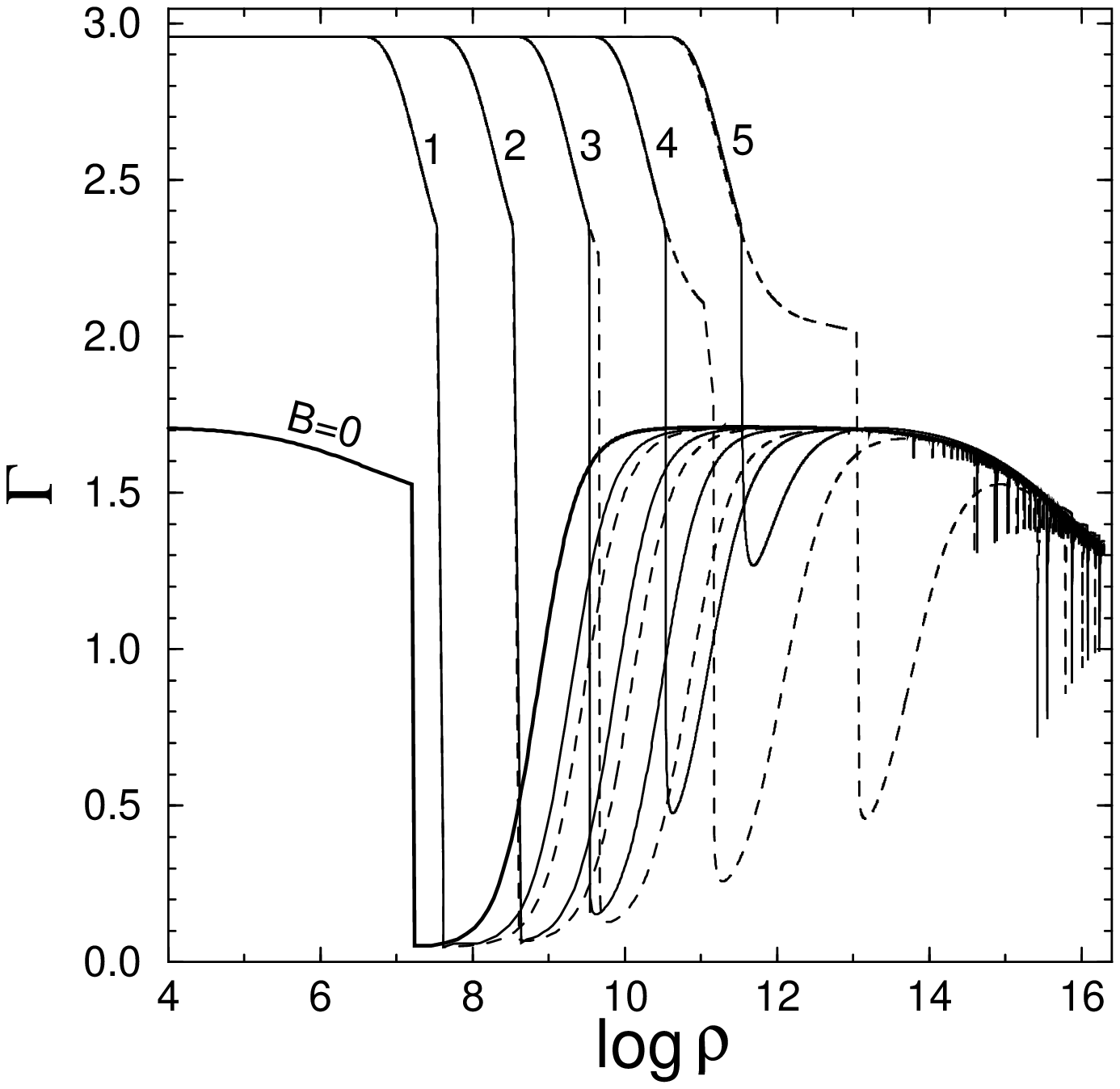}}
\vskip 0.2cm
\end{center}
{\small {\sc Fig. 5}
Adiabatic index $\Gamma$ vs. $\rho$ (in ${\rm \gcm}$) for a $npe$ gas in
various magnetic fields. Curves are labeled as in Fig. 4.
}
\vskip 0.5cm
\section{Muon production in a strong magnetic field}

In order to study the appearance of new particles at high density,
let us consider muons in an ideal $npe$ gas. 
Normally, muons decay to electrons via,
\be
\mu^{-} \rightarrow e^{-} \; + \nu_{\mu} + \bar{\nu_{e}}. \label{muon1}
\ee
But when the Fermi energy of the electrons approaches the muon rest mass 
$\mmu \simeq 105.66$ MeV, it becomes energetically favorable for electrons 
at the top level of Fermi sea to decay into muons with neutrinos and antineutrinos 
escaping from the star.
Hence, above some density, muons and electrons are in equilibrium:
\be
\mu^{-} \leftrightarrow e^{-} \label{muon2}, 
\ee
assuming that the neutrinos leave the star.
In this chemical equilibrium, charge conservation implies
\be
\mu_{\mu} = \mu_{e} \label{muon3}.
\ee
Equilibrium between $n$, $p$, and $e$ means
\be
\mu_{n} = \mu_{p} + \mu_{e}  \label{muon4},
\ee
and charge neutrality requires
\be 
n_{p} = n_{e} + n_{\mu}. \label{muon5}
\ee

Now consider the minimum density at which muons are first produced in a strong magnetic 
field. The threshold condition for muons to appear is given as
$n_{\mu} = 0$. In order to satisfy this condition, $\epmu$ must be unity.
If $\epmu \ne 1$, the muon number density $n_{\mu}$ is not zero even though the maximum 
Landau level $n_{\max}^{\mu}$ for muons in a strong magnetic field is zero.  
Thus, from Eq. (\ref{muon3}), we simply obtain 
\be
\epe^{\ast} = \frac{\mmu}{\me}. \label{muon6}
\ee
Then $\epp^{\ast}$ and $\epn^{\ast}$ are given by chemical equilibrium, 
Eq. (\ref{muon4}), and the charge neutrality condition, Eq. (\ref{muon5}), 
for a given $\gme$ when $n_{\mu} = 0$. 
Actually, since $\epp^{\ast}$ and $\epn^{\ast}$ depend on the strength of the magnetic 
field, we solve Eqs. (\ref{muon4}) and (\ref{muon5}) simultaneously to obtain
\be
\epp^{\ast} = \sqrt{\frac{\mpr^2 + \mmu^2 - \me^2}{\mpr^2}} - \szp \kp \gmp,
\label{muon7}
\ee
and
\be
\epn^{\ast} = \frac{\mpr}{\mn} \epp^{\ast} + \frac{\me}{\mn} \epe^{\ast}. 
\label{muon8}
\ee  
Thus, the threshold density for the appearance of muons in 
a magnetic field becomes:
\be
\rho_{\mu}^{\ast} (B) = \rho_n (\epn^{\ast}) + \sum_{j=e,p} \frac{\gm_j}{4 \pi^2} 
\frac{m_j}{\lambda_{j}^3} \chi_{j} (\ep_{j}^{\ast}). 
\label{muon9}
\ee
Figure 6 shows the muon threshold density $\rho_{\mu}^{\ast} (B)$ as a function of $\gme$.
We can see that $\rho_{\mu}^{\ast} (B)$ without the AMM is not affected by magnetic fields 
less than $B \sim 10^{18}$ G. But when we consider the AMM, its contribution to 
$\rho_{\mu}^{\ast} (B)$ becomes important above $B \sim 10^{17}$ G. 
However, the equation of state for a $npe$ gas with or without a magnetic field
is nearly unaffected by the existence of muons in neutron stars (\cite{canuto}).

We should also correct for the appearance of hyperons (\cite{BJ}). Indeed light hyperons
with mass less than $\sim 1.2$ GeV are expected to appear at typical neutron star densities
depending on 
the model. But the resulting equation of state is not very different from that of
pure neutron matter (\cite{ST}).
\placefigure{fig6}
\begin{center}
\vspace*{0.5cm}
{\epsfxsize=6.5cm
\epsfbox{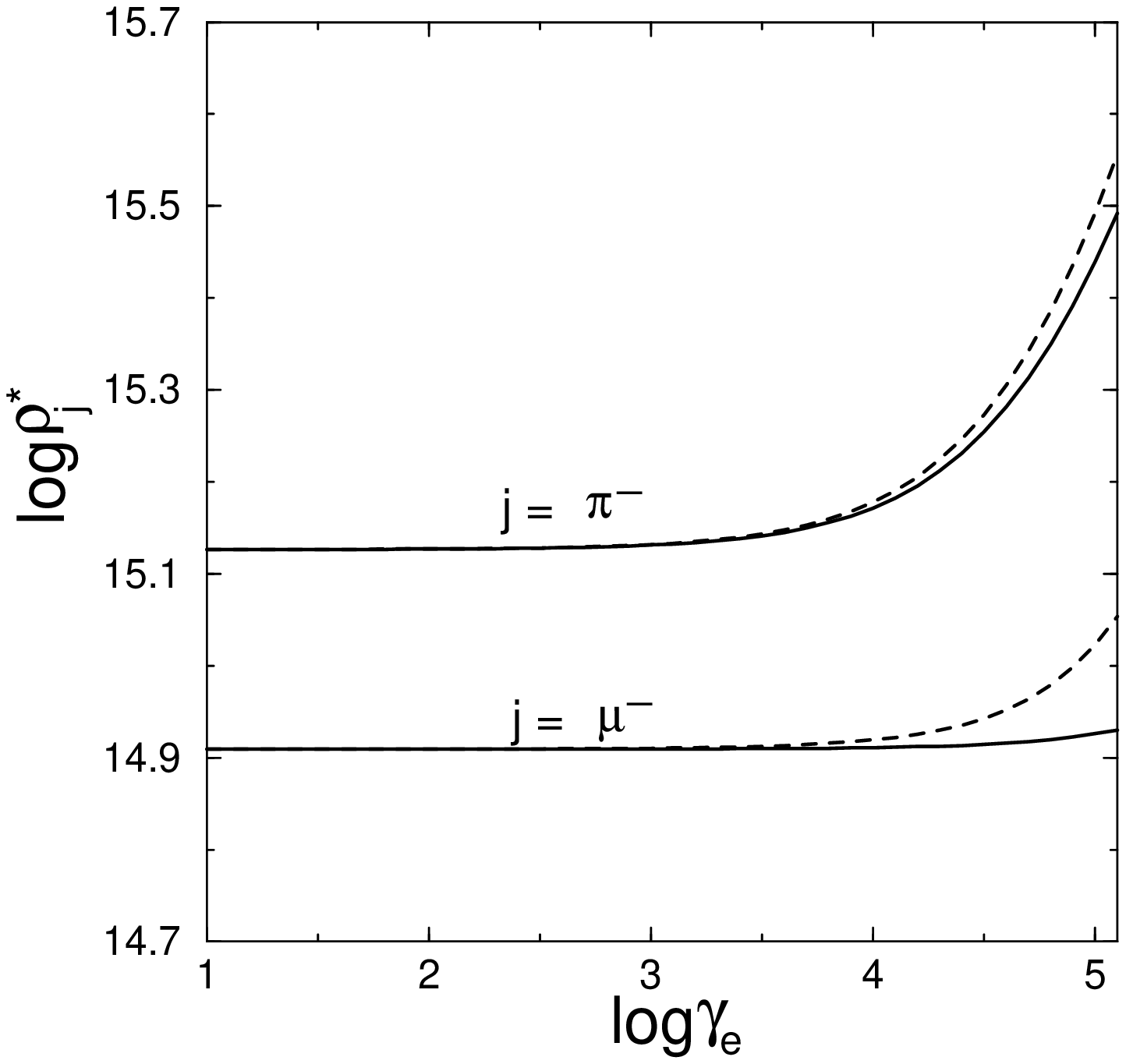}}
\vskip 0.2cm
\end{center}
{\small {\sc Fig. 6}
Appearance density $\rho_{j}^{\ast}$ (in ${\rm \gcm}$) for muons and pions
($j = \mu^{-}, \pi^{-}$) as a function of $\gme$. 
The solid line are for the cases without the AMM.
The dashed lines are the cases with the AMM included.
}
\vskip 0.5cm
\section{Pion production and condensation in a strong magnetic field}

At very high density ($\rho \gsim \rho_{n}^{\ast}$), neutron-rich nuclear matter
is no longer the true ground state of neutron-star matter. It will quickly
decay by weak interactions into chemically equilibrated neutron-star
matter. Fundamental constituents, besides neutrons, may then include a fraction of protons, 
hyperons, and possibly more massive baryons. 
Moreover, a phase transition to quark matter and boson (pion, kaon) condensation are also
possible. However, a first-order phase transition by thermal nucleation to quark matter 
may not occur in a magnetic field (\cite{chakrabarty95}).
Boson condensation in an external magnetic field is also a very subtle problem.

Recently, a true Bose-Einstein condensation (BEC) has been experimently realized in 
a system of $^{87}$Rb atoms that was confined by magnetic fields and evaporatively 
cooled (\cite{anderson}).  
An important consequence of the possible appearance of spin zero bosons is that
they can form a Bose-Einstein condensation (BEC) at sufficiently low temperatures.
An ideal condensation consists of a large number of bosons in a state of zero
kinetic energy. This would have at least two implications. 
One is that the equation of state would be softened, and the other is that the cooling rate 
from escaping neutrinos would be enhanced (\cite{ST}).  

Regarding the BEC in a magnetic field, long ago Schafroth (1955) pointed out that 
for a non-relativistic boson gas, BEC does not take place in the presence of a constant 
magnetic field. It also was shown by Toms (1994, 1995) that generally 
a BEC in the presence of a constant magnetic field does not occur in any number of 
spatial dimensions. However, recently Elmfors et al. (1995) suggested
that condensation may occurs in three dimension since the Landau ground state can
accommodate a large charge density even though it is not exactly a BEC.
In particular, Rojas (1996) has shown that a BEC actually may occur in the presence of
a constant homogeneous magnetic field in three dimension without requiring the vanishing
of the chemical potential. There is, however, no critical temperature at which
condensation begins.

Actually, since the criterion for condensation (usually taken as the equality of the
chemical potential to the ground state energy) leads to a divergence of the number
density, condensation cannot occur in a magnetic field. 
However, if the chemical potential depends on both temperature and the
magnetic field the divergence is avoided and condensation may occur (\cite{rojas}). 
Eventually, in a strong magnetic field and at any non-zero temperature, 
the number density in the ground state becomes finite.  
Therefore, in this work we will assume that boson condensation occurs in a strongly 
magnetized neutron star and that the boson number density has a finite value at sufficiently 
low temperature. 

In order to obtain the energy dispersion relation for a spin zero charged boson,
we solve the Klein-Gordon equation in an external magnetic field.
Under the same conditions as for charged fermions of Section 2, 
the dispersion relation for a charged boson in a magnetic field is then given by
\be
E_{\nb}^b = [p_{z}^2 c^2+ m_{b}^2 c^4+ \hbar c e B \nb]^{1/2}, 
\label{pion1}
\ee  
where $\nb \equiv 2 n + 1, \, (n = 0, 1, \, \ldots$),
$n$ is the principal quantum number of the Landau level, and $b$ denotes 
charged bosons ($b = \pi^{\pm}, \, K^{\pm}, \cdots$). 
Note that for charged bosons in a magnetic field, their energy state 
depends on the magnetic field strength even though they are in the lowest Landau level 
($n=0$). Similarly, the boson state density in the absence of a magnetic field,
\be
\frac{1}{\hbar^3} \int \frac{d^{3}\vec{p}}{(2 \pi)^3},   
\label{pion2}
\ee
is now replaced with
\be
\frac{1}{\hbar^2 c} \sum_{n=0}^{\infty} \int {eB\over(2\pi)^2} dp_z 
\label{pion3}
\ee
in a magnetic field.
 
Neutron stars will provide a unique opportunity to verify the hypothesis of boson
(pion, kaon) condensation in a strong magnetic field.
In this work, as an example, we only consider charged pion condensation via 
$n \rightarrow p + \pi^{-}$. 
Since an ideal cold $npe\mu$ gas allows $\pi^{-}$'s to be produced,
let us consider the pion appearance threshold in a strongly magnetized neutron star.
If we neglect the strong interaction between pions and nucleons, negatively charged pions
are formed through the reaction:
\be
n \longrightarrow p + \pi^- , 
\label{pion4}  
\ee
in dense neutron matter 
when the electron chemical potential $\mu_e$ exceeds the $\pi^-$ rest mass, 
$\mpi = 139.6$ MeV. 
Chemical equilibrium requires that the chemical potential should satisfy 
\be
\mu_{n} - \mu_{p} = \mu_{e} = \mu_{\pi} \;\;\; \rm{and} \;\;\; \mu_{e} = \mu_{\mu}
\label{pion5}
\ee
Charge neutrality also requires
\be
n_e + n_{\mu} + n_{\pi} = n_p , 
\label{pion6}
\ee
that is, 
\ba
\sum_{j = e, \mu} \Bigg[ 
\frac{\gm_j}{2 \pi^2 \lambda_{j}^3} \sum_{n = 0}^{n_{max}^j} \sum_{s_{z}^j}
\sqrt{(\ep_j - \kappa_j)^2 - (1 + 2 \gm_j n_{f}^j)} \Bigg] + n_{\pi} \nn \\
= \frac{\gmp}{2 \pi^2 \lamp^3} \sum_{n = 0}^{n_{max}^p} \sum_{\szp}
\sqrt{\epp^2 - \mpt^2}. 
\label{pion7} 
\ea
Therefore, we can obtain the total mass-energy density of a $npe\mu\pi$ gas:
\be
\rho = \sum_{j=n,p,e,\mu} \rho_j + \mpi n_{\pi}. 
\label{pion8}
\ee
For a given $\rho$, we can determine all the quantities
$\epe, \, \epmu, \, \epp, \, \epn$, and $n_{\pi}$ from Eqs. (\ref{pion5}) - (\ref{pion8}).
Figure 7 shows the pion number density $n_{\pi}$ as a function of $\rho$ for a given 
magnetic field. For ${\rm log}\gme \leq 3$, the magnetic field effect is negligible.
\placefigure{fig7}
\begin{center}
\vspace*{0.5cm}
{\epsfxsize=6.5cm
\epsfbox{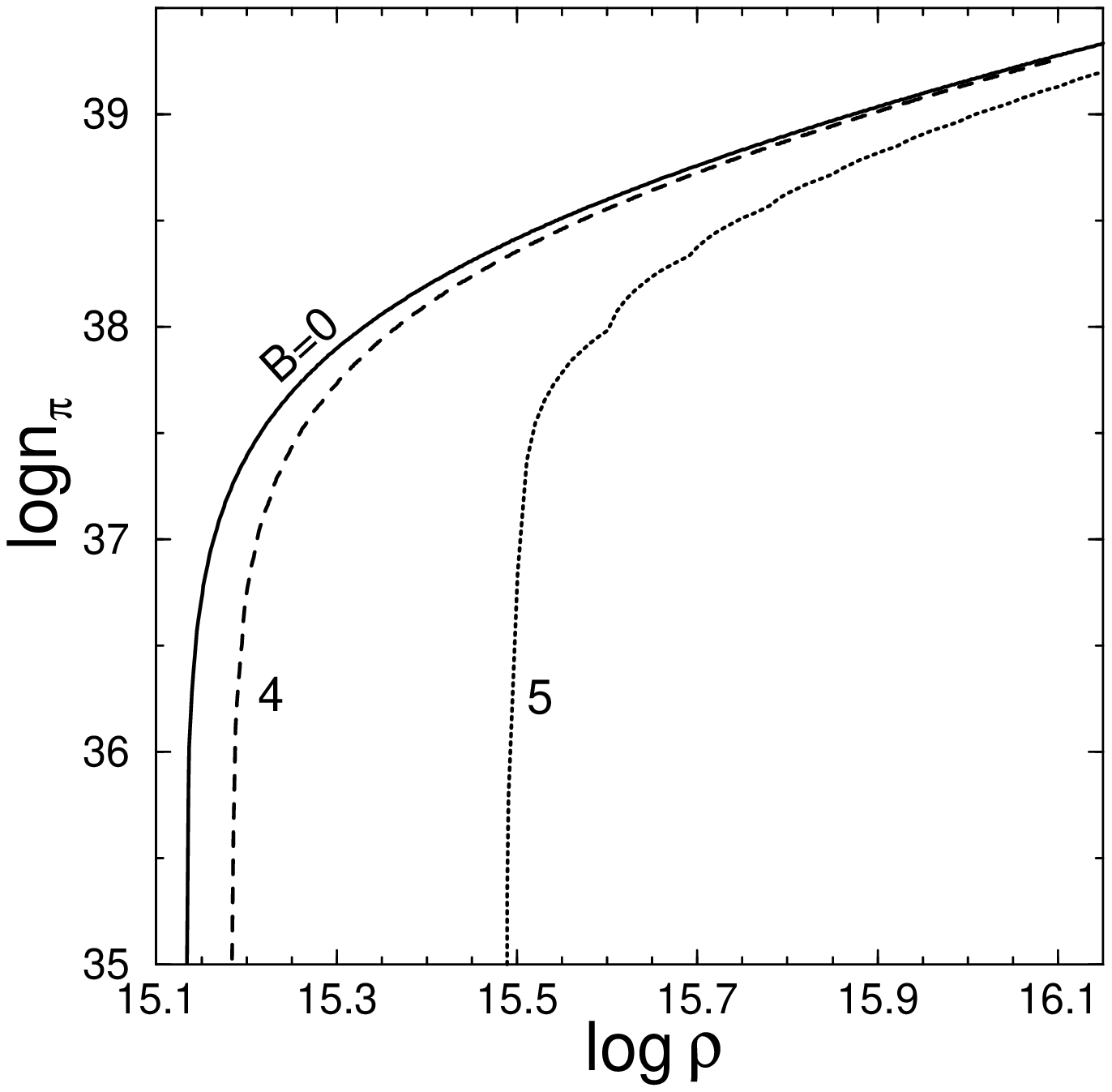}}
\vskip 0.2cm
\end{center}
{\small {\sc Fig. 7}
The pion number density $n_{\pi}$ (in ${\rm cm^{-3}}$) as a function of $\rho$
(in ${\rm \gcm}$)
for the log$\gme = 4$ (dashed line) and log$\gme = 5$ (dotted line).
The solid line denotes the non-magnetic case (B=0).
}
\vskip 0.5cm
Similar to the muon appearance, we see that the threshold condition for $\pi^{-}$ 
production is given as  $n_{\pi} = 0$.
Finally, the pion production density in a magnetic field is given by 
\be
\rho_{\pi}^{\ast} (B) = \rho_n (\epn^{\ast}) + \sum_{j=e,p,\mu} \frac{\gm_j}{4 \pi^2}
\frac{m_j}{\lambda_{j}^3} \chi_{j} (\ep_{j}^{\ast}). 
\label{pion9}
\ee
Figure 6 also shows the pion threshold density $\rho_{\pi}^{\ast} (B)$ as a function of $\gme$.
We also can see that $\rho_{\pi}^{\ast} (B)$ is not affected by magnetic fields less than
$B \sim 10^{17}$ G. However, above this field strength, the magnetic field effect on 
$\rho_{\pi}^{\ast} (B)$ is important. Note that since pions appear at the extremely high density
of $\rho \sim 10^{15}$ ${\rm \gcm}$, the AMM effect is small. 

From Eq. (\ref{pion5}), the pion momentum is zero ($\xpi = 0$) because of the fact that in the
ground state the condensed pions have zero kinetic energy. 
This forces the electron chemical potential $\epe$ to remain constant 
for $\rho > \rho_{\pi}^{\ast}$.
In consequence, the electron number density and pressure remain constant as $\rho$ increases.
Hence, increasing the pion density contributes to the total mass-energy density 
but not the pressure.
As a result, for a given $\rho$, the pressure in the condensate phase is always
less than in the non-condensate phase. 
Figure 8 shows the equation of state for 
an ideal magnetic $npe\mu\pi$ gas with pion condensation. We can see that magnetic fields
reduce the pion condensation. However, we still have a distinguishable pion condensate
equation of state in strongly magnetized neutron stars. However, at this high density,
the AMM does not contribute to the equation of state for the magnetized $npe\mu\pi$ gas
although at $\gme \gsim 10^5$ the nucleon AMM changes somewhat the pion number density.
For ${\rm log}\gme \leq 3$, the magnetic field effect is negligible and nearly the same
as the non-magnetic $npe\mu\pi$ gas.
Figure 9 shows the adiabatic index $\Gamma$ as a function of $\rho$ for a pion condensate
equation of state. Here we also can see the oscillatory behavior of the adiabatic index. 
The inclusion of the AMM makes a secondary oscillation in the adiabatic index, 
but this secondary oscillation disappears gradually as density increases. 

\section{Discussion}

In this work, we have studied the nuclear equation of state for an ideal $npe$ gas
in a strong magnetic field. In particular, we have calculated the proton concentration,
the threshold densities for neutron, muon, and pion production and pion condensation
in a strong magnetic field without and with the effect of the nucleon anomalous magnetic 
moments. 
In these calculations, we have shown that the higher Landau levels are significant 
at high density in spite of the existence of a very strong magnetic field.
In particular, at high density, the proton concentration approaches the nonmagnetic limit.
As a result, inverse $\beta$ decay is still suppressed in intense magnetic fields.
Therefore, neutron-star rapid cooling is probably not affected by the direct URCA process 
which is enhanced in strong magnetic fields.
In particular, we have obtained the neutron appearance density in a magnetic field when
the nucleon AMM is included. 
We also show that the muon and pion appearance density are not affected by magnetic fields 
less than about $B \sim 10^{17}$ G.
Finally, we here obtained an equation of state with a pion condensate in strong magnetic fields.
Magnetic fields reduce the amount of pion condensation.
However, we still have distinguishable effects of a pion condensate
equation of state in strongly magnetized neutron stars.
In addition, we found the oscillatory behavior of the adiabatic index in both strongly magnetized
$npe$ and $npe\mu\pi$ gases at high density. 
Here we speculate that this behavior may also lead to an interior pulsational instability.

It is generally accepted that neutrons and protons in a $npe$ gas are superfluid.
The charged pion condensate is also superfluid and superconductive (\cite{migdal}).
This pion formation and condensation in dense nuclear matter would have the significant
consequence (\cite{SM20c}) that the equation of state would be softened. 
First of all, softening the equation of state reduces the maximum mass of the stars
(\cite{baym}).
This softening effect with pion condensation also leads to detectable predictions 
(\cite{migdal}).
These are: (i) the enhanced rate of neutron star cooling via neutrinos.
(ii) a possible phase transition of neutron stars to a superdense state; and
(iii)  sudden glitches in pulsar periods.
In particular, if the pion condensation occurs in a strong magnetic field, 
it may significantly affect starquakes.

According to the magnetar model by Duncan and Thompson (1992, 1995), SGRs are caused by
starquakes in the outer solid crust of the magnetar. As a colossal magnetic field shifts,
it strains the crust with huge magnetic forces and sometimes it cracks.
When the crust snaps, it vibrates with seismic waves similar to those of an earthquake.
However, in neutron stars they also produce a flash of soft gamma-rays.
In addition, Cheng and Dai (1998) recently suggested that SGRs may be rapidly rotating
magnetized strange stars with superconducting cores.

Although such models can explain some crucial features, there are still several
unsettled issues (\cite{liang}). Therefore, superconducting cores with a charged boson
(pion and/or kaon) condensate in magnetars might be an alternative model to explain the
energy source of soft gamma-rays from magnetars.
\placefigure{fig8}
\begin{center}
{\epsfxsize=6.5cm
\epsfbox{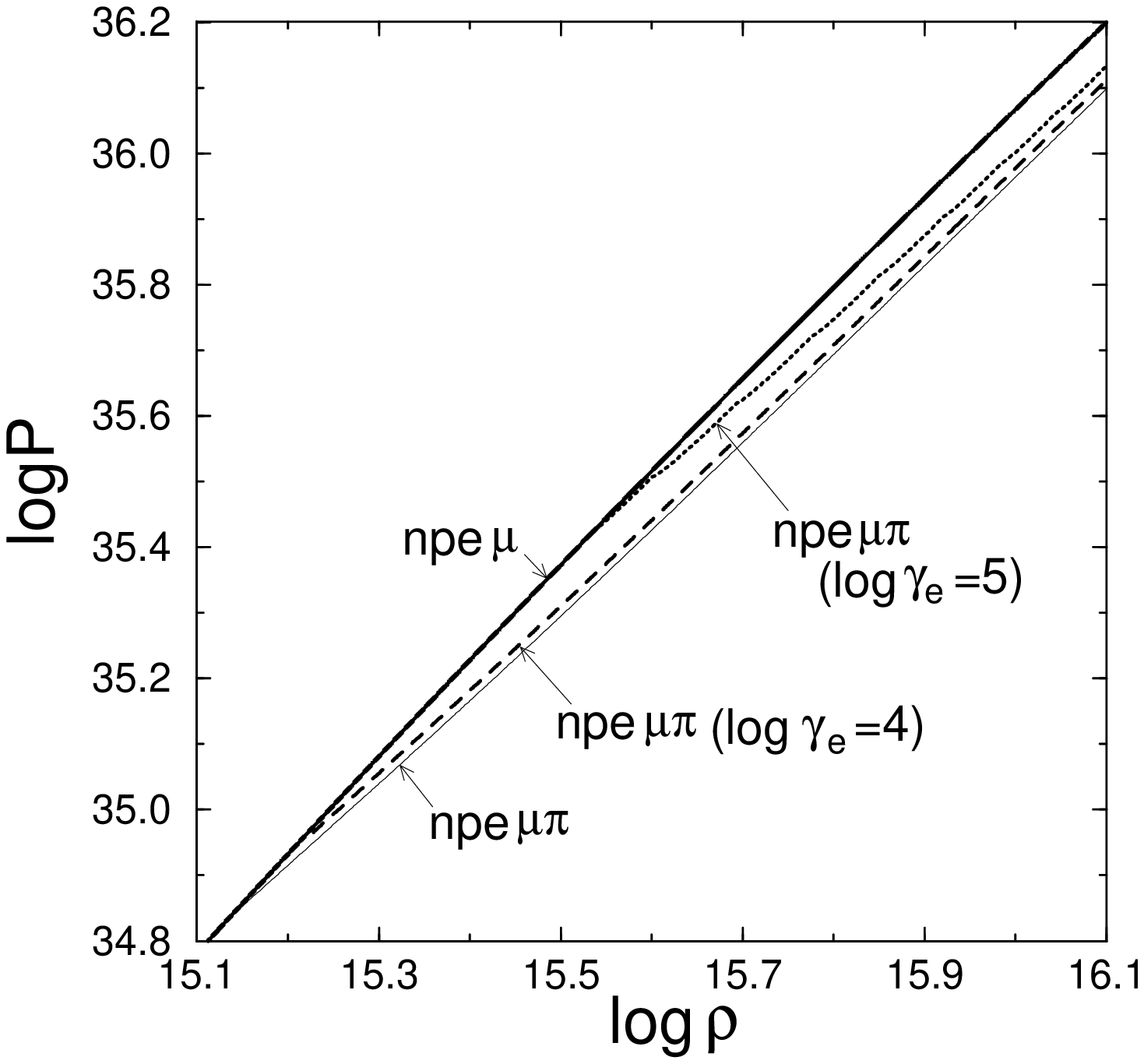}}
\vskip 0.2cm
\end{center}
{\small {\sc Fig. 8}
The equation of state for an ideal magnetic $npe\mu\pi$ gas with pion condensation.
The thick and thin solid lines are non-magnetic cases ($B=0$).
The dashed and dotted lines are magnetic cases for log$\gme = 4$ and $5$,
respectively.    
}
\vskip 0.5cm
\placefigure{fig9}
\begin{center}
{\epsfxsize=6.5cm
\epsfbox{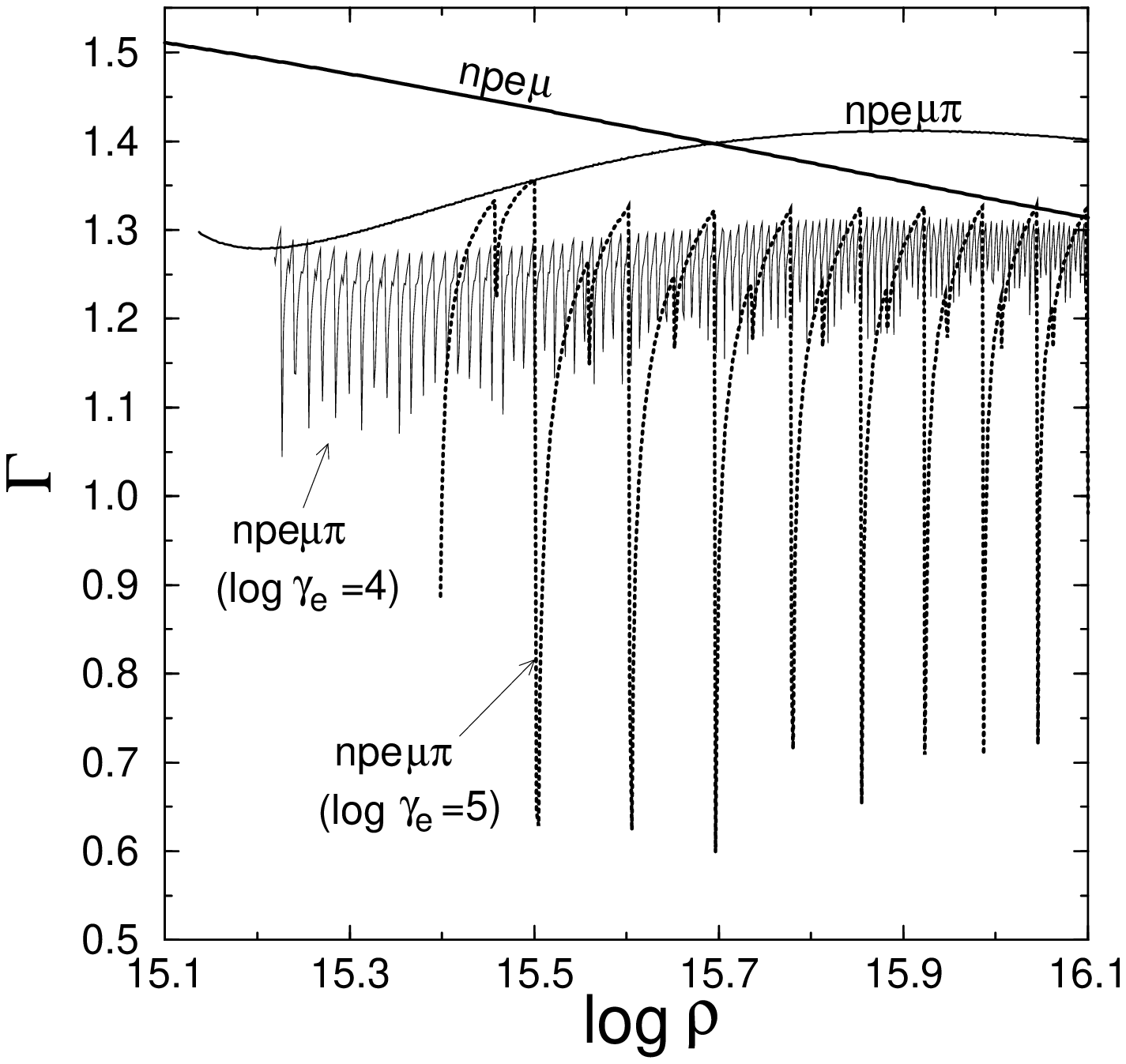}}
\vskip 0.2cm
\end{center}
{\small {\sc Fig. 9}
Adiabatic index $\Gamma$ as a function of $\rho$ (in ${\rm \gcm}$) for a pion condensate
equation of state for magnetic field strengths ${\rm log}\gme = 4$ and $5$.
The thick and thin solid lines are for non-magnetic ($B=0$) $npe\mu$ and $npe\mu\pi$ cases.
}
\vskip 0.5cm

\acknowledgments 
The authors are grateful to an anonymous referee for a careful reading of the paper and
his kind replies.
We also would like to thank J. M. Lattimer, F. Weber, and J. R. Wilson for 
stimulating comments and discussions. 
This work supported in part by DOE Nuclear Theory Grant DE-FG02-95ER40934.


\end{document}